\documentclass[jkps,twoside,twocolumn,fleqn,showpacs,showkeys]{revtex4}
\usepackage{graphicx}
\usepackage{amssymb}
\usepackage{amsmath}
\usepackage{bm}
\begin{document}
\setcounter{page}{0}
\title[]{Electric Dipolar Susceptibility of the Anderson-Holstein Model}
\author{Takahiro \surname{Fuse}}
\email{fuse-takahiro@tmu.ac.jp}
\author{Takashi \surname{Hotta}}
\affiliation{Department of Physics,
Tokyo Metropolitan University,
Hachioji, Tokyo 192-0397, Japan}


\begin{abstract}
The temperature dependence of electric dipolar susceptibility $\chi_P$
is discussed on the basis of the Anderson-Holstein model
with the use of a numerical renormalization group (NRG) technique.
Note that $\chi_P$ is related with phonon Green's function $D$.
In order to obtain correct temperature dependence of $\chi_P$
at low temperatures,
we propose a method to evaluate $\chi_P$ through the Dyson equation
from charge susceptibility $\chi_c$ calculated by the NRG,
in contrast to the direct NRG calculation of $D$.
We find that the irreducible charge susceptibility estimated from
$\chi_c$ agree with the perturbation calculation,
suggesting that our method works well.
\end{abstract}





\maketitle

In the research field of condensed matter physics,
exotic magnetism in cage structure materials such as filled skutterudites
has attracted much attention due to the interests on new electronic properties
caused by oscillation of a guest atom in a cage composed of
relatively light atoms \cite{Rattling}.
Such oscillation with large amplitude is frequently called {\it rattling}
and it is considered to play crucial roles for the formation of
magnetically-robust heavy electron state
in SmOs$_4$Sb$_{12}$ \cite{SmOs4Sb12}.
This peculiar heavy-electron state has been theoretically investigated
from various aspects by several groups
\cite{Miyake1,Hattori1,Hattori2,Mitsumoto1,Mitsumoto2,Mitsumoto3,
Fuse1,Fuse2,Fuse3,Hotta1,Hotta2,Hotta3,Hotta4,Hotta5,Hotta6}.

Recently, the Kondo effect of a {\it vibrating} magnetic ion in a cage
has been theoretically discussed on the basis of
a two-channel conduction electron system
hybridized with a vibrating magnetic ion
\cite{Yashiki1,Yashiki2,Yashiki3}.
Note here that a vibrating ion inevitably
induces electric dipole moment.
Then, it has been found that magnetic and non-magnetic Kondo effects
alternatively occur due to the screening of spin moment and
electric dipole moment of vibrating ion \cite{Hotta7}.
In particular, electric dipolar two-channel Kondo effect has been
found to occur for weak Coulomb interaction.
Then, it has been proposed that
magnetically robust heavy-electron state appears
near the fixed point of electric dipolar two-channel Kondo effect.

In this paper, in order to promote our understanding on
the Kondo effect concerning electric dipole moment $P$,
we analyze the temperature dependence of electric dipolar
susceptibility $\chi_P$
on the basis of the Anderson-Holstein Hamiltonian
with the use of a numerical renormalization group method.
For the reproduction of correct temperature
dependence of $\chi_P$ at low temperatures,
we propose a method to evaluate $\chi_P$ through the Dyson equation
from charge susceptibility $\chi_c$.
This method is found to provide correct results
in the temperature region lower than the Kondo temperature,
in sharp contrast to the numerical evaluation
of the phonon Green's function
which is directly related to $\chi_P$.

Now we explain the model Hamiltonian.
We consider a conduction electron system
in which an impurity ion is embedded.
On the impurity site, localized electrons are coupled with
ion vibration.
The situation is well described by the Anderson-Holstein model,
given by \cite{unit}
\begin{equation}
  \label{eq.AHmodel}
  H\!=\!\sum_{\bm{k}\sigma} \varepsilon_{\bm{k}}c_{\bm{k}\sigma}^\dag c_{\bm{k}\sigma}
   \!+\! V \sum_{\bm{k}\sigma}(c_{\bm{k}\sigma}^\dag d_\sigma +{\rm h.c.})\!+\!H_{\rm loc},
\end{equation}
where $\varepsilon_{\bm{k}}$ is the dispersion of conduction electron,
$c_{\bm{k}\sigma}$ is an annihilation operator of conduction electron
with momentum $\bm{k}$ and spin $\sigma$,
$d_\sigma$ denotes an annihilation operator of the localized electron
with spin $\sigma$,
and $V$ is the hybridization between conduction and localized electrons. 
The local term $H_{\rm loc}$ is given by 
\begin{equation}
H_{\rm loc}=\mu\rho +gx\rho +p^2/2 +\omega^2 x^2 /2,
\end{equation}
where $\mu$ is a chemical potential,
$\rho=\sum_\sigma n_\sigma$ with $n_\sigma=d_\sigma^\dag d_\sigma$,
$g$ denotes the coupling between electron density and ion vibration,
$x$ is normal coordinate of the vibrating ion,
$p$ indicates the corresponding canonical momentum,
and $\omega$ is the vibration energy.
Note that the reduced mass of the vibrating ion is set as unity.

In the present model, we ignore the Coulomb interaction term
$U n_\uparrow n_\downarrow$.
Of course, the effect of $U$ is quite important,
but our main purpose here is to understand
the temperature dependence of electric dipolar susceptibility $\chi_P$.
Since $\chi_P$ should be affected by $U$,
it is necessary to grasp in the first place
the temperature dependence of $\chi_P$ without
the effect of $U$.
Thus, in the present work, the Coulomb interaction is ignored.
The effect of $U$ will be discussed elsewhere in future.

For actual calculations, it is convenient to introduce
phonon operator $b$ and $b^{\dag}$
through the relation of $x=(b+b^\dag)/\sqrt{2\omega}$.
Then, the local term is rewritten as
\begin{equation}
 H_{\rm loc}= \mu\rho +
 \sqrt{\alpha}\omega\rho(b+b^\dag)+\omega(b^\dag b+1/2), 
\end{equation}
where $\alpha$ is the non-dimensional electron-phonon coupling,
defined by $\alpha=g^2/(2\omega^3)$.
Concerning the average electron number, throughout this paper,
we consider the half-filling case, at which $\mu$
is given by $\mu=2\alpha \omega$.

The conduction electron model hybridized with local impurity
is precisely analyzed with the use of a numerical renormalization group
(NRG) technique \cite{NRGWilson,NRGKrishna}.
The logarithmic discretization of the momentum space
is characterized by a parameter $\Lambda$
and we keep $M$ low-energy states
for each renormalization step.
Throughout this paper,
Here we set $\Lambda$=2.5 and $M$=5000.
The energy unit is a half-bandwidth of the conduction electron.
In this unit, we fix $V$ as $V=0.25$.
As for the calculation of phonon part,
the number of phonon basis is $50$.

In order to clarify electronic properties of $H$,
first we evaluate entropy $S_{\rm imp}$ and
specific heat $C_{\rm imp}$,
which are given by 
$S_{\rm imp}$=$-\partial F/\partial T$
and 
$C_{\rm imp}$=$-T\partial^2 F/\partial T^2$,
respectively, where $F$ is the free energy of
local electron and $T$ is a temperature,
defined by $T=\Lambda^{-(N-1)/2}$ with
the renormalization step number $N$.
We also evaluate charge and spin susceptibilities,
$\chi_c$ and $\chi_s$, which are, respectively, given by
\begin{equation}
 \chi_c=\frac{1}{Z}\sum_{i,j}\frac{e^{-E_i/T}-e^{-E_j/T}}{E_j-E_i}
\left|\langle i|\rho-\langle \rho \rangle |j\rangle\right|^2,
\end{equation}
and
\begin{equation}
 \chi_s=\frac{1}{Z}\sum_{ij}\frac{e^{-E_i/T}-e^{-E_j/T}}{E_j-E_i}
\left|\langle i| s_z |j\rangle\right|^2,
\end{equation}
where $E_i$ is the eigen-energy of $H$,
$|i\rangle$ is the corresponding eigen-state,
$Z$ is the partition function given by $Z=\sum_i e^{-E_i/T}$, 
$\langle \rho \rangle = \sum_i \langle i | \rho | i \rangle e^{-E_i/T}/Z$,
and $s_z=n_\uparrow - n_\downarrow$.

\begin{figure}
\includegraphics[width=65mm]{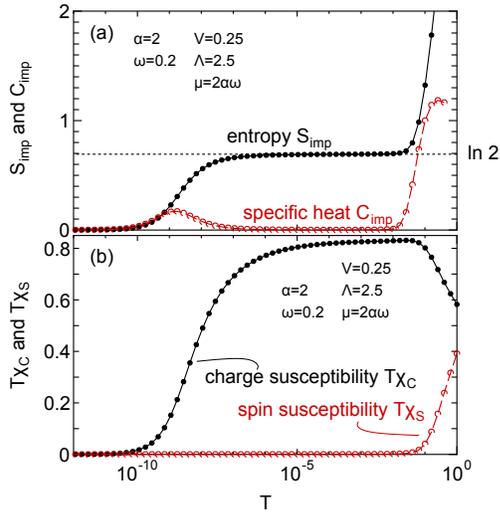}
\caption{(Color online)
(a) Entropy $S_{\rm imp}$ and specific heat $C_{\rm imp}$
vs. temperature $T$ for $\alpha=2$ and $\omega=0.2$.
(b) Charge susceptibility $T\chi_c$ and spin susceptibility $T\chi_s$
vs. temperature $T$ for the same parameters as (a).
}\label{Fig1}
\end{figure}

Now we move on to the NRG results.
First we show typical results for the Kondo phenomenon
due to the electron-phonon coupling.
In Fig.~\ref{Fig1}(a), we depict $S_{\rm imp}$ and $C_{\rm imp}$
as functions of $T$ for $\alpha=2$ and $\omega=0.2$.
In the high-$T$ region as $T \gtrsim 10^{-2}$,
$S_{\rm imp}$ shows a large value due to phonon excited states.
For $10^{-7} \lesssim T \lesssim 10^{-2}$,
$S_{\rm imp}$ indicates a plateau of $\ln 2$
and $C_{\rm imp}$ becomes almost zero in the corresponding
temperature region.
Note that we do not understand what degree of freedom is
relevant to $\ln 2$ only from the results of
$S_{\rm imp}$ and $C_{\rm imp}$.
This point will be discussed later.
Around at $T=1.8\times 10^{-9}$, we observe the release of
entropy $\ln 2$ and the specific heat $C_{\rm imp}$ forms
a clear peak, which defines the Kondo temperature $T_{\rm K}$.
Thus, we obtain $T_{\rm K} =1.8\times 10^{-9}$ for
the present parameters.
Finally, for $T \lesssim 10^{-9}$,
both $S_{\rm imp}$ and $C_{\rm imp}$ vanish,
indicating that the system is
in the local Fermi-liquid state.

In order to clarify the degree of freedom relevant to
the present Kondo phenomenon,
we evaluate charge and spin susceptibilities.
The results are shown in Fig.~\ref{Fig1}(b).
We immediately notice that
the spin susceptibility $\chi_s$ is rapidly suppressed even
at high temperatures as $T \sim10^{-1}$,
when we decrease the temperature.
On the other hand, the charge susceptibility $\chi_c$ is rather
increased to the enhanced value for $T\sim 10^{-2}$.
In the temperature range
of $10^{-7} \lesssim T \lesssim 10^{-2}$
in which $\ln 2$ plateau appears,
we find the large value of $\chi_c$.
Around at $T=T_{\rm K}$, we observe that
$\chi_c$ is gradually suppressed.
Namely, the Kondo behavior clearly appears in
the temperature dependence of $\chi_c$.
We conclude that in the present model,
the charge Kondo effect occurs,
since degenerate vacant and double occupied states
play roles of pseudo spins.

As for the understanding of the charge Kondo effect,
it seems to be enough to evaluate $\chi_c$.
However, in order to visualize the situation of the charge Kondo effect,
it is useful to recall that the vibrating ion induces
electric dipole moment $P$, given by $P=zex$,
where $z$ denotes the valence of the guest ion,
$e$ indicates electron charge, and $x$ is ion displacement.
Namely, the electric dipole susceptibility $\chi_P$ is also related to
the Kondo effect for the vibrating ion problem \cite{Hotta7}.
As easily understood from the definition of $P$, we obtain $\chi_P$ as
\begin{equation}
\chi_P =\frac{(ze)^2}{2\omega}D,
\end{equation}
where $D$ denotes the zero-energy component of
the phonon Green's function, given by
\begin{eqnarray}
 D_{\rm NRG}=\frac{1}{Z}\sum_{ij}\frac{e^{-E_i/T}-e^{-E_j/T}}{E_j-E_i}
 \left| \langle i | u-\langle u \rangle | j \rangle  \right|^2.
\label{eq.Ddrc}
\end{eqnarray}
Here $u$ is given by $u=b+b^{\dag}$ and
we add a subscription ``NRG'' to show explicitly that this quantity
is evaluated by the NRG method.

\begin{figure}
\includegraphics[width=70mm]{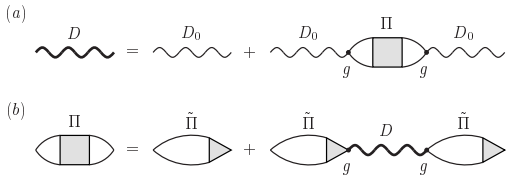}
\caption{
Dyson equations for phonon Green's function $D$ (thick wavy line)
and polarization function $\Pi$ (bubble with hatched square).
Thin wavy line and bubble with hatched triangle denote
non-interacting phonon Green's function $D_0$
and irreducible polarization function ${\tilde \Pi}$,
respectively.
}\label{diag}
\end{figure}

Of course, we can perform the NRG calculation for $D_{\rm NRG}$,
but we consider an alternative way to evaluate $\chi_P$ without using $D_{\rm NRG}$.
For the purpose, we exploit the Dyson equation,
which relates $\chi_c$ and $D$.
In general, the Dyson equations are diagrammatically
shown in Fig.~\ref{diag} and the first one (a) is expressed by
\begin{equation}
  D(i\nu_n) = D_0(i\nu_n) + g^2 {D_0(i\nu_n)}^2 \Pi(i\nu_n),
\end{equation}
where $\nu_n=2\pi T n$ is the boson Matsubara frequency with an integer $n$,
$D$ is the dressed phonon Green's function,
$g$ is the electron-phonon coupling constant,
$\Pi$ denotes polarization function,
and $D_0$ is a non-interacting phonon Green's function,
given by $D_0(i\nu_n)=-2\omega/[(i\nu_n)^2-\omega^2]$.
At the static limit of $\nu_n=0$,
by noting that $\Pi(0)=\chi_c$,
$D_0(0)=2/\omega$, and $g=\sqrt{\alpha}\omega$,
we obtain a relation between $D$ and $\chi_c$ as
\begin{equation}
  D = 2/\omega + 2 \alpha \chi_c.
  \label{eq.Ddys}
\end{equation}
In principle, $D$ of eq.~(\ref{eq.Ddys}) is equal to 
$D_{\rm NRG}$ of eq.~(\ref{eq.Ddrc}).

\begin{figure}
\includegraphics[width=70mm]{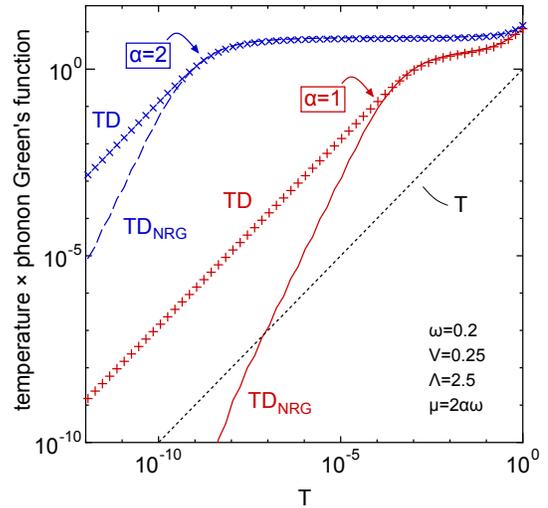}
\caption{ (Color online)
$D$ and $D_{\rm NRG}$ vs. temperature $T$ for $\omega=0.2$.
The solid (dashed) curve indicates $TD_{\rm NRG}$ for $\alpha=1$ ($2$),
while the cross (plus) indicates $TD$ for $\alpha=1$ ($2$).
}\label{Fig2}
\end{figure}

In Fig.~\ref{Fig2}, we depict $TD_{\rm NRG}$ and $TD$ vs. $T$ for $\omega=0.2$
for the comparison of $D_{\rm NRG}$ and $D$.
We show a couple of results for $\alpha=1$ and $2$.
In the high-$T$ region, we find that $TD_{\rm NRG}$ and $TD$ agree well
with each other.
To be honest, we observe a little quantitative difference between
$TD_{\rm NRG}$ and $TD$, although it is difficult
to notice it in the graph of the logarithmic scale.
This type of small deviation can be overcome by the
elevation of the numerical accuracy.
For instance, if we keep more numbers of states
in each renormalization step,
the accuracy is expected to be improved.

However, in the low-temperature region of $T < T_{\rm K}$,
where $T_{\rm K} = 1.8\times 10^{-9}$ for $\alpha=2$
and $T_{\rm K} =6.6\times 10^{-4}$ for $\alpha=1$,
we observe serious discrepancy between $TD_{\rm NRG}$ and $TD$.
From the analysis of numerical data,
we find that $TD_{\rm NRG} \propto T^2$ and $TD \propto T$
for $T < T_{\rm K}$.
When we fix a temperature and increase $\alpha$,
both $TD_{\rm NRG}$ and $TD$ are monotonic increasing functions.
The slope of $TD_{\rm NRG}$ or $TD$ seems to be
independent to $\alpha$.

This difference seems to be so serious that it is difficult to
determine which is the correct behavior.
However, here we recall the local Fermi-liquid theory
for $T < T_{\rm K}$.
Namely, we obtain the narrow quasi-particle band at the Fermi level
in the electron density of states,
leading to constant density of states $\rho_0$.
Since $\chi_c$ can be related to $\rho_0$ at low temperatures,
it is natural to consider that $D$ should be constant at the
same temperature region.
If we further imagine that phonon frequency ${\tilde \omega}$ is
decreased due to the Kondo screening,
the dressed phonon Green's function $D=2/{\tilde \omega}$
should be larger than $D_0(0)=2/\omega$.
We deduce that ${\tilde \omega}/\omega$ is a monotonic decreasing
function of $\alpha$.
In short, it seems to be natural to consider
that $TD \propto T$ in the charge Kondo effect.

If so, here we have a naive question concerning the reason why
we cannot evaluate correctly $D$ by the NRG method.
In eq.~(\ref{eq.Ddrc}), $D_{\rm NRG}$ can be calculated in the NRG method,
but it includes implicit problems in the calculation accuracy.
In the NRG method, in order to deal with the states which increase rapidly
in each renormalization step,
only $M$ low-energy states are kept and the rests are simply discarded.
In the renormalization process,
the electron-phonon excited states which are
indispensable for the description of phonon excitation will be lost.
This type of problem has been already pointed out in the analysis of
the two-channel model \cite{Hotta7}.
On the other hand, 
since the local charge susceptibility $\chi_c$ is the physical quantity
with large contribution from the electron states near the Fermi level,
we expect that $\chi_c$ is obtained in good accuracy with the use of NRG method.

Let us check the NRG calculation from a different viewpoint
by the comparison with the perturbation calculation.
Here we define the irreducible polarization function ${\tilde \Pi}(i\nu_n)$,
which is related with $\Pi$ and $D$ through the Dyson equation (b)
in Fig.~\ref{diag} as
\begin{equation}
  \Pi(i\nu_n) = {\tilde \Pi}(i\nu_n) + g^2 {\tilde \Pi}(i\nu_n)^2 D(i\nu_n).
\end{equation}
At $\nu_n=0$, by noting eq.~(\ref{eq.Ddys}) and $\chi_c=\Pi(0)$,
we obtain
\begin{equation}
 {\tilde \chi}_c^{-1}=\chi_c^{-1} +2 \alpha \omega,
 \label{irrchic}
\end{equation}
where irreducible charge susceptibility is defined as
${\tilde \chi}_c$=${\tilde \Pi}(0)$.

\begin{figure}
\includegraphics[width=60mm]{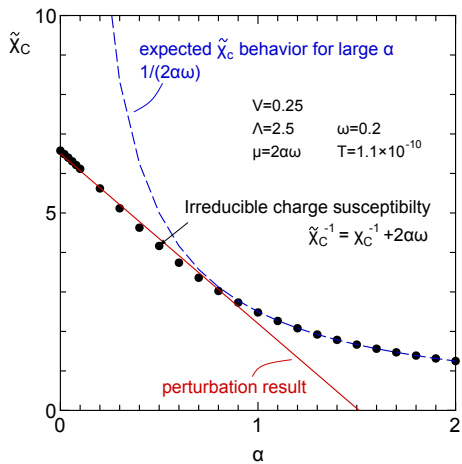}
\caption{ (Color online)
Irreducible charge susceptibility $\tilde{\chi}_c$ (solid circles)
vs. $\alpha$ for $T=1.1\times 10^{-10}$ and $\omega=0.2$.
The solid line indicates the perturbation calculation result of ${\tilde \chi}_c$
up to the first order of $\alpha$ and the dashed curve means the expected behavior
of ${\tilde \chi}_c \sim 1/(2\alpha\omega)$ for large $\alpha$, respectively. 
}\label{Fig3}
\end{figure}

In Fig.~\ref{Fig3}, we show ${\tilde \chi}_c$ vs. $\alpha$ obtained from eq.~(\ref{irrchic})
with the use of the NRG result of $\chi_c$ for $\omega=0.2$.
Note that $\chi_c$ at low enough temperature is estimated at $T=1.1\times 10^{-10}$.
For each value of $\alpha$ , we have checked that
${\tilde \chi}_c$ is unchanged even at lower $T$.
For small $\alpha$, it is shown that the NRG results agree quite well with
the perturbation result in the order of $\alpha$.
For large $\alpha$, $D$ is enhanced and $\chi_c$ becomes large,
indicating that ${\tilde \chi}_c$ approaches the value of $1/U_{\rm ph}$,
where $U_{\rm ph}$ denotes the magnitude of attractive interaction,
given by $U_{\rm ph}=2 \alpha \omega$ in the present case.
The NRG results for ${\tilde \chi}_c$ reproduce the behavior of $1/U_{\rm ph}$
for large $\alpha$.
Thus, it is confirmed that $\chi_c$ is constant at low temperature from
eq.~(\ref{irrchic}).

Finally, let us provide a comment on the effect of Coulomb interaction,
which has been completely ignored in this paper.
When we include $U$, we imagine that $U_{\rm ph}$ is suppressed
as $U_{\rm ph}=2\alpha\omega-U$.
For $U>2\alpha \omega$, the charge Kondo effect disappears and
the spin Kondo effect occurs instead \cite{Hotta3}.
We think that it is interesting to consider the behavior of $\chi_P$
for $U \ge 2\alpha \omega$.

In summary, we have discussed the temperature dependence of
electric dipolar susceptibility $\chi_P$ of the Anderson-Holstein model
with the use of the NRG technique.
We consider a direct method to evaluate phonon Green's function $D_{\rm NRG}$
and another indirect way to calculate $D$ from $\chi_c$ through the
Dyson equation.
After careful investigations, we have concluded that
$D$ provides correct temperature dependence in contrast to $D_{\rm NRG}$
for $T$ smaller than the Kondo temperature.
The effect of the Coulomb interaction on $\chi_P$
is one of future problems.

\begin{acknowledgments}

We thank Kazuo Ueda for useful discussions.
This work has been supported by Grant-in-Aids
for Scientific Research on Innovative Areas ``Heavy Electrons''
(No.~20102008) for the Ministry of Education, Culture,
Sports, Science, and Technology, Japan,
and for Scientific Research (C) (No.~24540379)
from Japan Society for the Promotion of Science.
The computation in this work has been partly done using the facilities
of the Supercomputer Center of Institute for Solid State Physics,
University of Tokyo.

\end{acknowledgments}

\end{document}